\documentclass[prd,superscriptaddress,nofootinbib,floatfix,11pt]{revtex4-2} 
\usepackage[utf8]{inputenc}
\usepackage{amsfonts,amsmath,amssymb}
\usepackage{slashed}
\usepackage{graphicx}
\usepackage{subfigure}
\usepackage[colorlinks=true]{hyperref}
\usepackage[usenames,dvipsnames]{color}
 \usepackage[normalem]{ulem}
\usepackage{geometry}
\usepackage{bm}               
\usepackage{xspace}	
\usepackage{epstopdf}
\usepackage{pgffor}					
\usepackage{longtable,booktabs}
\usepackage{mathtools}
\usepackage{subfigure}

\newcommand{\all}{\text{all}}

\usepackage{newtxtext, newtxmath}
\usepackage{braket,bm}
\usepackage{multirow}
\usepackage{enumitem}

\newcommand{\itp}{\affiliation{CAS Key Laboratory of Theoretical Physics, Institute of Theoretical Physics,\\ Chinese Academy of Sciences, Beijing 100190, China}}

\newcommand{\ucas}{\affiliation{School of Physical Sciences, University of Chinese Academy of Sciences, Beijing 100049, China}}

\newcommand{\uestc}{\affiliation{School of Physics, University of Electronic Science and
Technology of China,\\ Chengdu 610054, China}} 

\begin{document}

\title{$D^+D^-$ hadronic atom and its production in $pp$ and $p\bar{p}$ collisions  }

\author{Pan-Pan Shi}\email{shipanpan@itp.ac.cn}
\author{Zhen-Hua Zhang}\email{zhangzhenhua@itp.ac.cn}
\author{Feng-Kun Guo} \email{fkguo@itp.ac.cn}
\itp \ucas

\author{Zhi Yang}\email{zhiyang@uestc.edu.cn} \uestc

\begin{abstract}

There must be Coulomb bound states of a pair of hadrons, which are stable against the strong interaction, with opposite electric charges. Such bound states are hadronic atoms.
We study the properties and the production of the ground-state $D^+D^-$ hadronic atom $A_{D^+D^-}$, called dionium, with quantum numbers $J^{PC}=0^{++}$. 
Using a nonrelativistic effective field theory for the $D^0\bar{D}^0$-$D^+D^-$ coupled-channel system, the mass of the ground-state dionium is predicted to be $(3739.3 \pm 0.1)~\text{MeV}$, with the binding energy reduced by about 10\% compared to the Coulomb binding energy due to the strong interaction. Its width for the decay into the neutral $D^0\bar D^0$ channel is predicted to be $1.8^{+1.4}_{-0.6}$~keV using lattice inputs for the $D\bar D$ strong interaction. 
The cross section for the inclusive prompt production of the dionium at CMS and LHCb and that for the direct production $p\bar p\to A_{D^+D^-}$ at PANDA are estimated at an order-of-magnitude level. 
In particular, we expect that 
 $\mathcal{O}(10^3\sim 10^5)$ events of the reaction chain $p\bar p\to A_{D^+D^-}\to D^0\bar D^0 \to K^-\pi^+K^+\pi^-$ can be collected at PANDA, and valuable information on the charmed meson interaction and on understanding charmoniumlike states will be obtained. 
 
\end{abstract}

\date{\today}

\maketitle

\newpage
\section{Introduction}

Since 2003, many charmoniumlike states have been reported at various experiments, e.g. the $X(3872)$ and $Z_c(3900)^\pm$ with quantum numbers $J^{PC}=1^{++}$ and $J^{PC}=1^{+-}$, respectively~\cite{Belle:2003nnu,BESIII:2013ris,Belle:2013yex,BESIII:2013qmu}, which, being close to the $D\bar D^*$ thresholds, are good candidates of hadronic molecules (composite systems of hadrons bound together through the strong interaction) although there are other interpretations in the market (for recent reviews, see Refs.~\cite{Chen:2016qju,Hosaka:2016pey,Lebed:2016hpi,Esposito:2016noz,Guo:2017jvc,Olsen:2017bmm,Liu:2019zoy,Brambilla:2019esw,Guo:2019twa}) . 
To understand these charmoniumlike structures, it is important to study the interactions between a pair of charmed and anticharmed mesons. 

By making use of the approximate heavy quark spin symmetry (HQSS), spin partners of the $X(3872)$ as a $D\bar D^*$ hadronic molecule have been predicted~\cite{Nieves:2012tt,Hidalgo-Duque:2012rqv,Guo:2013sya} (see also Refs.~\cite{Tornqvist:1993ng,Wong:2003xk,Swanson:2005tn,Dong:2021juy}).
In particular, an isoscalar $D^*{\bar D}^*$ bound state with $J^{PC}=2^{++}$ was predicted to be just below the $D^*\bar D^*$ threshold with a width of the order of a few MeV to tens of MeV~\cite{Albaladejo:2015dsa,Baru:2016iwj}. The prediction recently got support from the resonant structure observed by Belle in $\gamma\psi(2S)$~\cite{Belle:2021nuv}, which has a mass of $(4014.4\pm4.1\pm0.5)$~MeV, though the global significance is only 2.8$\sigma$.

Recently, a lattice QCD (LQCD) calculation of the $D\bar{D}$ and $D_s\bar{D}_s$ scattering in the energy region from slightly below 3.73 up to 4.13~GeV indicates a possible $D\bar{D}$ bound state with $J^{PC}=0^{++}$ and a binding energy of $E_B \equiv 2m_D - M=4.0^{+5.0}_{-3.7}$~MeV with $m_D(M)$ the mass of the $D$ meson (the bound state), which has also been predicted by several phenomenology models~\cite{Wong:2003xk,Gamermann:2006nm,Nieves:2012tt,Hidalgo-Duque:2012rqv,Zhang:2006ix,Liu:2009qhy,Dong:2021juy}.  Various methods have been proposed to search for this state in experiments~\cite{Dai:2015bcc,Dai:2020yfu,Wang:2020elp,Xiao:2012iq}.

Regardless of whether the isoscalar $D\bar{D}$ bound state exists, a $D^+D^-$ hadronic atom (composite system of hadrons bound together mainly through the electric Coulomb interaction), to be called dionium and denoted as $A_{D^+D^-}$ in the following, must exist and may be found in the $D^0\bar{D}^0$ invariant mass distribution from some high-energy reactions. It is a composite state of the $D^+$ and $D^-$ mesons formed mainly through the Coulomb interaction. The Bohr radius of the dionium is 
\begin{equation}
    r_B=\frac{1}{\alpha\mu_c}=(6.82~\text{MeV})^{-1}=28.91~\text{fm},
    \label{eq:rb}
\end{equation}
with $\alpha={1}/{137}$ the fine structure constant and $\mu_c$ the reduced mass of the $D^+D^-$ mesons, much larger than $\Lambda_{\text{QCD}}^{-1}\sim \mathcal{O}(1/300~\text{MeV}^{-1})$, the typical scale of strong interaction. Therefore, the strong interaction between $D^+$ and $D^-$ can be treated as higher-order corrections in the dionium, and the nonrelativistic effective field theory (NREFT) can be applied to the dionium. 
Hadronic atoms formed by light hadrons have been extensively studied in the NREFT framework in order to measure the scattering lengths of light-hadron scattering; for reviews, see Refs.~\cite{Gasser:2007zt,Gasser:2009wf}. In the charmonium energy region, a hadronic atom formed by charmed mesons $D^{\pm}$ and $D^{*\mp}$, called the $X$ atom, is proposed and investigated in Ref.~\cite{Zhang:2020mpi}; its binding energy is $E_{B_A}=22.92$~keV, and the production rates in $B$ decays
and prompt $pp$ collisions relative to the those of the $X(3872)$ are predicted, in a parameter-free way, to be $\gtrsim 1\times 10^{-3}$, where the suppression is due to the Coulomb character of the wave function.
An experimental search of the $X$ atom can provide crucial information toward resolving the debates regarding the nature and production mechanism of the $X(3872)$.

In this work, we study the $D^0{\bar D}^0$-$D^+D^-$ coupled-channel scattering and derive the binding energy, decay width of the dionium, and coupling constants between the dionium and $D{\bar D}$ meson pairs. Due to the possible existence of an isoscalar $D\bar{D}$ hadronic molecular state mentioned above, the strong interaction between the $D\bar{D}$ should be treated nonperturbatively, different from the pionic atoms~\cite{Gasser:2007zt,Gasser:2009wf},\footnote{{There are also hadronic atoms in the light-hadron sector that involve nonperturbative strong interactions in the near-threshold region, such as the kaonium ($K^+K^-$ atom)~\cite{Dumbrajs:1985ay,Kerbikov:1995ge,Krewald:2003ab,Klevansky:2011hi} and the protonium ($p\bar p$ atom, for a review, see Ref.~\cite{Klempt:2002ap}). 
These systems are usually treated theoretically by solving the Schr\"odinger equation with both the Coulomb and strong interaction potentials.
Different from the dionium to be discussed, for these two systems, the neutral channels, $K^0\bar K^0$ for kaonium and $n\bar n$ for protonium, have thresholds higher than the charged ones and thus do not contribute to their decay widths.
}
} and the binding energy of the $D\bar{D}$ molecular bound state from LQCD can be used as an input to fix the low energy constant (LEC) in the NREFT.
Analogous to the $X$ atom, a search of the dionium may also provide useful information in understanding the heavy-meson interaction and thus the nature of the charmoniumlike states.
To search for the dionium, we estimate the cross sections for inclusive productions of the ground state, to be denoted as $A_{D^+D^-}$, at LHC, and for direct exclusive productions at PANDA. The inclusive production rate of this hadronic atom, $\sigma(pp\to A_{D^+D^-} +\all)$, at LHC is estimated by the following strategy: we first employ a Monte Carlo (MC) event generator, P{\scriptsize{YTHIA}}8~\cite{Sjostrand:2014zea}, to simulate the inclusive production of the $D^+D^-$ mesons in $pp$ collisions, and then the production of the dionium from the final-state
interaction (FSI) of the $D^+D^-$ pair is handled by NREFT. For the production of the dionium at PANDA through the process $p{\bar p}\rightarrow A_{D^+D^-}$,
we make use of the flavor $\text{SU}(4)$ model to estimate the $D^+D^-$ pair production rate, and the dionium production from the FSI is same as that in $pp$ collisions.  

This paper is organized as follows. In Sec.~\ref{Sect:pole_atom}, the properties of the dionium, 
including its binding energy, decay width, and coupling to $D{\bar D}$, are investigated. The production rates of the dionium in $pp$ collisions are discussed in Sec.~\ref{Sect:production_pp}.
The prediction for the $p{\bar p}\rightarrow A_{D^+D^-}$ cross section is given in Sec.~\ref{Sect:production_ppbar}. 
The results of our work are summarized in Sec.~\ref{Sect:summary}. Appendix~\ref{Sect:one_phase_space} contains the derivation of the production cross section of $p{\bar p}\rightarrow A_{D^+D^-}$.

\section{$D^+D^-$ hadronic atom}\label{Sect:pole_atom}

In this section, by considering the $D^0\bar{D}^0$-$D^+D^-$ coupled-channel system, we derive the binding energy and decay width of the dionium in terms of the $D\bar{D}$ scattering lengths, which is known as the Deser-Goldberger-Baumann-Thirring (DGBT) formula~\cite{Deser:1954vq,Gasser:2007zt,Gasser:2009wf} for hadronic atoms, and evaluate their numerical value and the couplings  between the dionium and $D{\bar D}$ mesons in the presence of an isoscalar $D\bar{D}$ hadronic molecular bound state.

We consider the $D^0\bar{D}^0$-$D^+D^-$ coupled-channel system because the dionium and the $D\bar D$ molecular bound state can be treated simultaneously in this framework. Both the $D\bar D$ hadronic molecule and the energy levels of the dionium will appear as poles of the coupled channel $T$-matrix, and the decay widths of the dionium to $D^0\bar{D}^0$ can be derived from the imaginary part of the dionium poles.

For the $D^0\bar{D}^0$-$D^+D^-$ coupled-channel system,
the thresholds of the two channels are denoted as $\Sigma_0=2m_{D^0}$ and $\Sigma_c=2m_{D^{\pm}}$, respectively, with $m_{D^{\pm}}$ the mass of the $D^{\pm}$ mesons and $m_{D^0}$ ($m_{\bar{D}^0}$) the mass of the $D^{0}$ ($\bar{D}^0$) meson, and the difference
between these two thresholds is $\Delta=\Sigma_c-\Sigma_0$. 
The initial energy relative to the $D^+D^-$ threshold is defined as $E=\sqrt{s}-\Sigma_c$, with $\sqrt{s}$ the total initial energy of the system.  
The scattering amplitude is given as~\cite{Braaten:2017kci}
\begin{align}
    \mathbf{T}(E)
    =\begin{pmatrix} 0 & 0 \\ 0 & T_c(E) \end{pmatrix}
    +\begin{pmatrix} 1 & 0 \\ 0 & W_c(E) \end{pmatrix} \mathbf{T}_s(E)
    \begin{pmatrix} 1 & 0 \\ 0 & W_c(E) \end{pmatrix} ,   
    \label{Eq:T_total}   
\end{align}
where $T_c(E)$ is the $S$-wave Coulomb 
scattering amplitude between $D^+$ and $D^-$,
\begin{align}
    T_c(E)=\frac{\pi}{\mu_c\kappa_c(E)}\left(\frac{\Gamma(1-\eta)}{\Gamma(1+\eta)}-1\right),
\end{align}
with $\kappa_c(E)=\sqrt{-2\mu_c E}$ and $\eta={\alpha\mu_c}/{\kappa_c(E)}$. 
Besides, $\mathbf{T}_s(E)$ contains the strong interaction contribution to $\mathbf{T}(E)$ as discussed in Ref.~\cite{Braaten:2017kci}, 
and $W_c(E)$ denotes the amplitude for the Coulomb photon exchange between the initial and final $D^+D^-$ states,
\begin{align}
    W_c(E)=\left(\frac{2\pi i \eta}{1-e^{-2\pi i\eta}}\frac{\Gamma(1-\eta)}{\Gamma(1+\eta)}\right)^{\frac{1}{2}}.
\end{align}

At leading order (LO), one can take a contact term~\cite{Hidalgo-Duque:2012rqv} as the potential of the strong interaction,
\begin{align}
    V_{\mathrm{LO}}=\frac{1}{2}\begin{pmatrix}
        C_{0a}+C_{1a} & C_{0a}-C_{1a}\\
        C_{0a}-C_{1a} & C_{0a}+C_{1a}
    \end{pmatrix},
\end{align}
where $C_{0a}$ and $C_{1a}$ denote the isoscalar and isovector interactions, respectively. 
Then, the $\mathbf{T}_s(E)$ in Eq. (\ref{Eq:T_total}) is given by the Lippmann-Schwinger equation (LSE),
\begin{equation}
    \mathbf{T}_s(E)=
    \mathbf{V}_{ \mathrm{LO}}^{\Lambda}+\mathbf{V}_{ \mathrm{LO}}^{\Lambda}\, \mathbf{G}^{\Lambda}(E)\, \mathbf{T}_s(E),
    \label{Eq:LSE}
\end{equation}
where $\mathbf{G}^{\Lambda}(E)$ is the Green's function regularized 
by a sharp cutoff $\Lambda$~\cite{Braaten:2017kci,Kong:1999sf,Konig:2015aka},
\begin{align}
    \mathbf{G}^{\Lambda}(E)=\begin{pmatrix}
        G_0^{\Lambda}(E) & 0\\
        0 &G_c^{\Lambda}(E)
    \end{pmatrix}.
    \label{Eq:DD_loop}
\end{align}
The Green's functions for the neutral  and charged channels are denoted as $G_0^{\Lambda}(E)$ and $G_c^{\Lambda}(E)$, respectively,
\begin{align}
    G_0^{\Lambda}(E)=&-\frac{\mu_0\Lambda}{\pi^2}+J_0(E),\label{Eq:G_0}\\
    G_c^{\Lambda}(E)=&-\frac{\mu_c\Lambda}{\pi^2}-\frac{\alpha\mu_c^2}{\pi}
        \left(\ln {\frac{\Lambda}{\alpha\mu_c}} -\gamma_E\right)+J_c(E),\label{Eq:G_c}
\end{align}
with
\begin{align}
    J_0(E)=&\,\frac{\mu_0}{2\pi}\sqrt{-2\mu_0(E+\Delta)},\\
    J_c(E)=&-\frac{\alpha\mu_c^2}{\pi}\left[\ln{\eta}+\frac{1}{2\eta}-\psi(-\eta)\right],
\end{align}
where $\mu_0$ is the reduced mass of $D^0{\bar D}^0$, $\psi(x)=\Gamma'(x)/\Gamma(x)$ is the digamma function, and $\gamma_E=0.577$ is the Euler constant.
The scattering amplitude $\mathbf{T}_s(E)$ can be solved from the LSE in Eq.~\eqref{Eq:LSE} as 
\begin{align}
    \mathbf{T}^{-1}_s(E)&=\left(\mathbf{V}_{\mathrm{LO}}^{\Lambda}\right)^{-1}-\mathbf{G}^{\Lambda}(E) \nonumber\\
    &=\frac{1}{2}\begin{pmatrix}
       \frac{1}{C_{0a}}+\frac{1}{C_{1a}} & \frac{1}{C_{0a}}-\frac{1}{C_{1a}}\\ \frac{1}{C_{0a}}-\frac{1}{C_{1a}} & \frac{1}{C_{0a}}+\frac{1}{C_{1a}} +\delta
    \end{pmatrix}
   -\begin{pmatrix}
        G_0^\Lambda(E)& \\ & G_c^\Lambda(E)
    \end{pmatrix} \\
    &=\begin{pmatrix}
        w_+-J_0(E) &-w_- \\ -w_-& w_+-J_c(E)+O\left(\frac{\alpha\mu_c}{\Lambda_{\mathrm{QCD}}}\right) 
    \end{pmatrix},
\end{align}
with
\begin{align}
    w_+=&\frac{1}{2C_{1a}^R}+\frac{1}{2C_{0a}^R},\qquad w_-=\frac{1}{2C_{1a}^R}-\frac{1}{2C_{0a}^R},\nonumber
\end{align}
where $C_{0a}^R$, $C_{1a}^R$ are the renormalized coupling constants, and an additional counterterm $\delta$ is added to count for the isospin breaking effect between the two channels
and to absorb the corresponding logarithmic UV divergence. After renormalization, the isospin breaking effect from the UV divergence gives a higher-order correction that can be neglected at LO.
Thus, the LO $\mathbf{T}_s(E)$ is 
\begin{align}
    \mathbf{T}_{s}(E)=\frac{1}{\mathrm{det}}
    \begin{pmatrix} 
          w_+-J_c(E) & w_- \\
         w_- & w_+-J_0(E)
    \end{pmatrix},
    \label{Eq:T_strong}
\end{align}
with 
\begin{align}
    \mathrm{det}=&\left[w_+-J_0(E)\right]\left[w_+-J_c(E)\right]
        -w_-^2.
    \label{Eq:w_det}
\end{align}

At the $D^0\bar{D}^{0}$ threshold, the $\mathbf{T}_s$ matrix elements can be written as
\begin{align}
    \mathbf{T}_{s00}(E=-\Delta)&\equiv-\frac{2\pi}{\mu_0}(a_0+a_1) =\frac{w_+-J_c(-\Delta)}{w_+\left[w_+-J_c(-\Delta)\right]-w_-^2} ,\nonumber\\
    \mathbf{T}_{s01}(E=-\Delta)&\equiv-\frac{2\pi}{\mu_0}(a_0-a_1) =\frac{w_-}{w_+\left[w_+-J_c(-\Delta)\right]-w_-^2},
       \label{Eq:scattering_lenght}
\end{align}
where $a_0$ and $a_1$ are the $D{\bar D}$ $S$-wave scattering lengths with isospin $I=0$ and $I=1$, respectively, and the renormalized coupling constants can be expressed in terms of the scattering lengths 
\begin{align}
    C_{1a}^R&\simeq\frac{-\frac{4\pi}{\mu_0}a_1}{1+\frac{2\pi}{\mu_c}(a_0-a_1)J_c(-\Delta)},\nonumber\\
    C_{0a}^R&\simeq\frac{-\frac{4\pi}{\mu_0}a_0}{1-\frac{2\pi}{\mu_c}(a_0-a_1)J_c(-\Delta)}.
\end{align}

The Coulomb binding energy of the $D^+D^-$ ground state is 
\begin{equation}
    E_1=\frac{\alpha^2\mu_c}{2}\simeq 24.9~\mathrm{keV}.
\end{equation}
The ground-state binding energy receives corrections from the strong interactions and will be different from the above value. The ground state is the lowest pole of $\mathbf{T}_s(E)$, and thus the binding energy $E_{A}$ satisfies
\begin{equation}
    \mathrm{det}|_{E=-E_{A}}=0.
    \label{Eq:pole_det}
\end{equation}
Separating the Coulomb pole of the ground state from the Coulomb propagator $J_c(E)$, it can be reexpressed as~\cite{Kong:1999sf} 
\begin{align}
    J'_c(E)=&-\frac{\alpha\mu_c^2}{\pi}\left[\ln\eta-\frac{1}{2\eta}-\frac{1}{\eta+1}+2-\psi(2-\eta)\right] +\frac{\alpha^3\mu_c^3}{\pi[E-(-E_1)]};     
\label{Eq:Jc_separate}
\end{align}
then, Eq.~\eqref{Eq:pole_det} is rewritten as
\begin{align}
   \frac{w_-^2}{w_+-J_0(-E_{A})} =w_+-J'_c(-E_A).
    \label{Eq:DDpole}
\end{align}
The LO ground energy level shift and the width of the decay into the $D^0\bar D^0$ channel can
be solved from Eq.~\eqref{Eq:DDpole} as 
\begin{align}
    \Delta E_1^{\mathrm{str}}-i\frac{\Gamma_1}{2}=&-E_{A}+E_1 \notag  \\
    =&\frac{\alpha^3\mu_c^3}{\pi}\frac{w_+(1-R) -i\frac{\mu_0}{2\pi}
    \sqrt{2\mu_0\Delta}R}{w_+^2(1-R)^2+\left(\frac{\mu_0}{2\pi}\right)^2 2\mu_0\Delta 
    R^2 } +\mathcal{O}\left(\frac{\alpha^5\mu_c^{5/2}}{\Delta^{3/2}}\right),
    \label{Eq:Energyshift}
\end{align}
which is the well-known DGBT formula~\cite{Deser:1954vq,Gasser:2007zt,Gasser:2009wf} for hadronic atoms, where $w_+$ and $w_-$ can be expressed in terms of the $D\bar{D}$ scattering lengths as 
\begin{align}
     w_+&=-\frac{\frac{\mu_0}{2\pi}(a_0+a_1)+(a_0-a_1)^2J_c(-\Delta)}{4a_0a_1},\\
    w_-&=-\frac{\frac{\mu_0}{2\pi}(a_0-a_1)+(a_0^2-a_1^2)J_c(-\Delta)}{4a_0a_1},
\end{align}
with
\begin{align}
    J_c(-\Delta)&=-\frac{\alpha\mu_c^2}{\pi}\left[\ln{\frac{\alpha\mu_c}{\sqrt{2\mu_c\Delta}}}+ \frac{\sqrt{2\mu_c\Delta}}{2\alpha\mu_c}-\psi \left(\frac{\alpha\mu_c}{\sqrt{2\mu_c\Delta}}\right)\right],
\end{align}
and the quantity $R$ in Eq.~\eqref{Eq:Energyshift} is defined as 
\begin{align}
   R\equiv \frac{w_-^2}{w_+^2+\left(\frac{\mu_0}{2\pi}\right)^2 2\mu_0\Delta}.
\end{align}
The real part of Eq.~\eqref{Eq:Energyshift} gives the LO strong energy level shift $\Delta E_1^{\mathrm{str}}$, 
and the LO decay width for the dionium decay to the neutral $D^0\bar{D}^0$ pair
is given by the imaginary part,
\begin{align}
    \Gamma_1=\frac{2\alpha^3\mu_c^3}{\pi}\frac{\frac{\mu_0}{2\pi}R\sqrt{2\mu_0\Delta}}{w_+^2(1-R)^2+\left(\frac{\mu_0}{2\pi}\right)^2 2\mu_0\Delta R^2 }.
\end{align}

In the presence of a shallow isoscalar $D\bar{D}$ bound state with $J^{PC}=0^{++}$ predicted in the LQCD  calculation~\cite{Prelovsek:2020eiw} and several phenomenological models (for a review, see Ref.~\cite{Dong:2021juy}), the energy level shift may receive further simplification. 
Since no isovector $D\bar{D}$ bound state is predicted, the contact potential for the isovector coupling is significantly weak as deduced from the properties of the $X(3872)$ with the help of HQSS, i.e., $|C_{0a}^{R}| \gg  |C_{1a}^{R}|$~\cite{Albaladejo:2015dsa}. We take $C_{1a}^{R}=0$, and then, at LO, the scattering matrix in Eq.~\eqref{Eq:T_strong} is reduced to
\begin{align}
    \mathbf{T}_s (E)=\frac{1}{\frac{2}{C_{0a}^{R}}-J_0(E)-J_c(E)}
    \begin{pmatrix}
        1 & 1\\
        1 & 1
    \end{pmatrix},
    \label{Eq:T_simp}
\end{align}
which is in line with the scattering amplitude in Ref.~\cite{Zhang:2020mpi}. The $D\bar{D}$ bound state appears as a pole of $\mathbf{T}_s(E)$ located at $E=-\Delta-E_B\equiv -\Delta_B$, with $E_B$ the binding
energy of the bound state, and the renormalized coupling constant $C_{0a}^{R}$ can be determined by
the pole location, 
\begin{equation}
    \frac{2}{C_{0a}^{R}}-J_0(E=-\Delta_B)-J_c(E= -\Delta_B)=0.
    \label{Eq:det_DDbar}
\end{equation}
The energy level shift in Eq.~\eqref{Eq:Energyshift} can be simplified as
\begin{align}
    \Delta E_1^{\text{str}}-i\frac{\Gamma_1}{2}&=-E_{A}+E_1 \nonumber\\
    &\simeq\frac{\alpha^3\mu_c^3}{\pi}\frac{\frac{\mu_0}{2\pi}\sqrt{2\mu_0E_B}+J_c(-\Delta_B)
    -i\frac{\mu_0}{2\pi}\sqrt{2\mu_0\Delta}}{\left[\frac{\mu_0}{2\pi}\sqrt{2\mu_0E_B}
    +J_c(-\Delta_B)\right]^2+\left(\frac{\mu_0}{2\pi}\right)^2 2\mu_0\Delta}.
\end{align}
The LQCD calculation in Ref.~\cite{Prelovsek:2020eiw} was performed without isospin symmetry breaking. Here,                    we take the $D\bar D$ threshold to be that of the $D^0\bar D^0$ pair to avoid  the situation that the bound state may be able to decay into the  $D^0\bar D^0$ channel within the LQCD uncertainties. Thus, setting $E_B=4.0^{+5.0}_{-3.7}~\mathrm{MeV}$~\cite{Prelovsek:2020eiw}, we can extract the binding energy and decay width of the ground-state dionium to the $D^0\bar{D}^0$ pair,
\begin{align}
    \text{Re}E_{A}&=E_1-\Delta E_1^{\mathrm{str}}\simeq 22.9_{-0.4}^{+0.3}~\mathrm{keV},\nonumber\\
    \Gamma_{1}&\simeq 1.8^{+1.4}_{-0.6}~\mathrm{keV},
    \label{Eq:Binding_energy}
\end{align}
where the uncertainties come from those of $E_B$. Note that the binding energy for the dionium is approximately equal to that of the $X$ atom, while their decay widths are largely different~\cite{Zhang:2020mpi} because of the $D^{*}$ decay width  involved in the $X$ atom.


The total decay width of the dionium receives additional contributions from channels other than the $D^0\bar D^0$, such as $J/\psi \pi\pi$ and purely-light-hadron channels. 
Such channels are expected to be subleading as the transitions from $D^+D^-$ to them would involve the exchange of a charmed meson. Thus, they are expected to be suppressed relative to the charge-exchange transition $D^+D^-\to D^0\bar D^0$, which may proceed through the exchange of light mesons (such as the $f_0$, $\rho$ and $\omega$). The $c\bar{c}$ pair annihilation is also orders of magnitude suppressed in comparison with that for a charmonium, as can be seen from the tiny ratio of the radial wave functions at the origin of the dionium ($R_{00}(0)=2\sqrt{\alpha^3\mu_c^3}\simeq 1\times10^{-3}$~GeV$^{3/2}$ for the ground state) to that of a charmonium (about 1~GeV$^{3/2}$ for the ground state~\cite{Eichten:1995ch}). 
Physically, this is because the  $c$ and $\bar{c}$ in the dionium are tens of fm apart from each other, see Eq.~\eqref{eq:rb}, a distance much larger than the typical scale of the strong interactions. 
Therefore, the partial width given in Eq.~\eqref{Eq:Binding_energy} is expected to give a good estimate of the total width of the ground state dionium.

The effective couplings of $D^+D^-$ to the ground-state dionium ($g_{\mathrm{str}}$) and $D^0\bar{D}^{0}$ to the
molecular bound state ($g_{0}$) can be extracted from the residues of $\mathbf{T}_{s22}$ and $\mathbf{T}_{s11}$,
\begin{align}
    g_{\mathrm{str}}^2&=\lim\limits_{E \to -E_{A}}(E+E_{A})\mathbf{T}_{s22}(E)\simeq \left[3.2_{-2.8}^{+2.1}-\left(3.5_{-1.5}^{+4.0}\right)i\right]\times 10^{-8}~\mathrm{MeV}^{-1} ,\nonumber\\
    g_{0}^2&=\lim\limits_{E \to -\Delta_{B}}(E+\Delta_{B})\mathbf{T}_{s11}(E)\simeq \left(3.9^{+1.4}_{-2.5}\right)\times 10^{-4}~\mathrm{MeV}^{-1} ,
    \label{Eq:coupling}
\end{align}
where the uncertainties come from the uncertainties of 
$E_B$, and the uncertainties of the charmed meson masses are negligible. 
Since $\left|g^2_{\text{str}}/g^2_0\right|\sim  O(10^{-4})$, the production rate for the dionium
is much smaller than that of the $D^0{\bar D}^0$ bound state with the same kinematics. 

\section{Production rate of the dionium in $pp$ collisions}\label{Sect:production_pp}

\begin{figure}[tbp]
    \includegraphics[width=0.45\columnwidth]{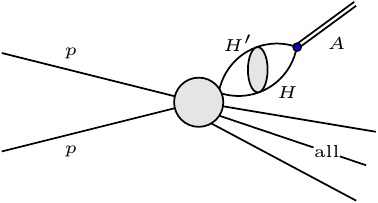}
    \caption{\label{Fig:pp_production}~The inclusive production of the hadronic atom made of the charged hadron pair $HH'$ in $pp$ collisions. The
    other hadrons produced in this process are generally denoted by ``$\all$.''} 
\end{figure}

The production of a hadronic atom $A$ in the $pp$ collision is shown in Fig.~\ref{Fig:pp_production}. The production of the $HH'$ pair is a short-distance process, and the hadronic atom arises from long- distance FSI of the $HH'$ pair. The scattering amplitude of this production is~\cite{Guo:2014sca,Guo:2014ppa}
\begin{align}
    {\cal M}[A+\all]={\cal M}[HH'+\text{all}]\times G \times T_A,
    \label{Eq:amp_pp}
\end{align}
where ${\cal M}[HH'+\text{all}]$ denotes the amplitude for the inclusive production of charged hadrons $H$ and $H'$ in the short-range interaction, 
$G$ is the Green's function of the $HH'$ pair~\cite{Gasser:2007zt,Zhang:2020mpi}, and $T_A$ is the amplitude for producing the hadronic atom in the process
$HH'\rightarrow A$. For a shallow bound $A$, $T_A$ is approximately equal to the coupling constant
 $g_{A}$ for  the hadronic atom $A$ to $HH'$ pair~\cite{Artoisenet:2009wk}. 

The general differential cross section  for the inclusive production of $HH'$ in the MC event generator is expressed as~\cite{Artoisenet:2009wk,Guo:2014sca}
\begin{align}
    d\sigma[HH'+\all]_{\text{MC}}=&\frac{1}{2}K_{HH'}\frac{1}{\text{flux}}\sum_{\text{all}}\int d\phi_{HH'+\text{all}} \left|{\cal M}[HH'+\text{all}]\right|^2\frac{d^3k}{(2\pi)^32\mu},
    \label{Eq:HHp_MC}
\end{align}
where $k$ is the relative 3-momentum of the $HH'$ pair in the center-of-mass frame, and $\mu$ is the reduced mass of the $HH'$ pair. 
Note that the factor $1/2$ is a Jacobian derived from coordinate transformation and the $K_{HH'}$ factor parametrizes the overall difference between the MC simulation and experimental data. In this work, 
we roughly take $K_{HH'}\simeq1$ for an order-of-magnitude estimate. 
The total cross section for the $A$ production 
in proton-proton collisions can be derived from Eqs.~\eqref{Eq:amp_pp} and \eqref{Eq:HHp_MC},
\begin{align}
    \sigma[A+\all]\simeq \frac{m_{A}}{m_{H}m_{H'}} \left|G g_{A}\right|^2\left(\frac{d\sigma[HH'+\all]}{dk}\right)_{\text{MC}}\frac{4\pi^2\mu}{k^2},
    \label{Eq:cross_section_pp}
\end{align}
 where $m_{H}$ and $m_{H'}$ are, respectively, the masses of hadrons $H$ and $H'$ 
 and $m_{A}$ is the mass of the hadronic atom $A$. 

For the production of the dionium, $HH'$ denotes the $D^+D^-$ pair. The Green's function for the $D^+D^-$ mesons is given by the loop function $G=G_c^{\Lambda}$, where $\Lambda$ is taken as a typical hadronic scale, $0.5\sim1.0$~GeV, for an estimate, and 
the coupling constant $g_A=g_{\text{str}}$; they are given in Eqs.~\eqref{Eq:G_c} 
and \eqref{Eq:coupling}, respectively. 

\begin{figure}[tbp]
    \includegraphics[width=0.7\columnwidth]{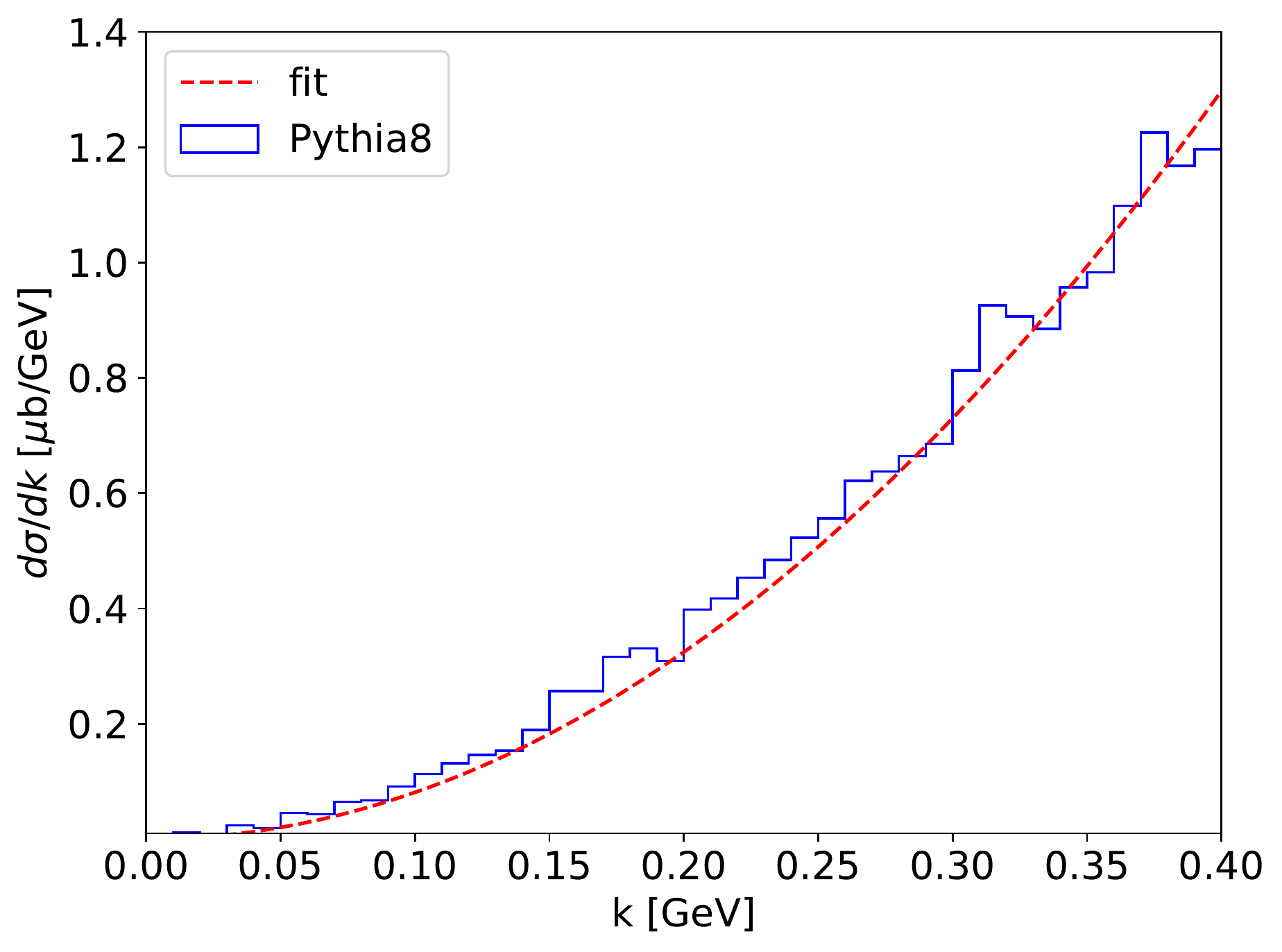}
    \caption{\label{Fig:pp_cross_section} ~The differential cross section $d\sigma/dk$ (in units of $\mu$b$/$GeV) 
    for the inclusive production of $D^+D^-$ in $pp$ collisions with $\sqrt{s}=7$~TeV. Here, $k$ is the relative 3-momentum of $D^+D^-$ pair.}
\end{figure}

The differential cross section for the production of $D^+D^-$ mesons at short distances can be estimated by the MC event generator, P{\scriptsize{YTHIA}}8~\cite{Sjostrand:2014zea}. 
Following the parameters setting at the CMS in the $X(3872)$ searching process~\cite{CMS:2013fpt}, 
 we generate partonic events in a full $2\rightarrow 2$ QCD process with a transverse momentum $10<p_T<50$ GeV and then hadronize these events to produce hadrons in $pp$ collisions. Consequently, we simulate $4\times 10^{9}$ $pp$ collisions with $pp$ center-of-mass energy $\sqrt{s_{pp}}=7$~TeV and select the $D^+D^-$ pair with a relative 3-momentum $k<0.4$~GeV. The obtained differential cross section  $d\sigma/dk$ (in units of $\mu$b$/$GeV) 
for the inclusive production of $D^+D^-$ pairs is shown in Fig.~\ref{Fig:pp_cross_section}. 
Considering a constant production amplitude, which is a good approximation in the small $k$ region, the differential cross section is proportional to $k^2$. Thus, fitting the differential cross section generated using P{\scriptsize{YTHIA}}8, such an expectation is confirmed, and one has
 \begin{align}
      \left(\frac{d\sigma[D^+D^-]}{dk}\right)_{\text{MC}} \simeq 3.34\times 10^6\frac{(k/\mu_c)^2}{s_{pp}(m_{D^+}+m_{D^-})}.
      \label{diff_cross_section_pp}
\end{align}

Finally, the integrated cross sections
for $pp\rightarrow A_{D^+D^-}+\all$ with different cutoffs are listed in Table~\ref{Tab:cross_section_pp}.
The prompt cross section for the production of the dionium at LHC is about $1(49)$~pb 
with the sharp cutoff $\Lambda=0.5(1.0)$ GeV. Comparing the cross section for $pp\rightarrow A_{D^+D^-}+\all$ 
and that of $pp\rightarrow X(3872)+\all$ predicted in Ref.~\cite{Guo:2014sca} at $\sqrt{s_{pp}}=7$~TeV, $\mathcal{O}(10~\text{nb})$, the predicted production rate for $A_{D^+D^-}$ relative to $X(3872)$ is at the $10^{-3}$ level, which is the same order of magnitude as the ratio of the production rates for the $X$ atom and the $X(3872)$~\cite{Zhang:2020mpi}. 
Therefore, at $\sqrt{s_{pp}}=7$~TeV,  the cross section of $pp\rightarrow A_{D^+D^-}+\all$ and $pp\rightarrow X_{\text{atom}}+\all$ are at the same order of magnitude, as one would naturally expect.

\begin{table}[tb]
\caption{\label{Tab:cross_section_pp} Order-of-magnitude estimates of the integrated cross section for $pp\rightarrow A_{D^+D^-}+\all$ at $\sqrt{s_{pp}}=7$~TeV with $10<p_T<50$~GeV.  
In the first row, $\Lambda$ is the sharp cutoff used to regularize the Green's function in Eq.~\eqref{Eq:G_c}.}
\renewcommand{\arraystretch}{1.2}
\begin{tabular*}{\columnwidth}{@{\extracolsep\fill}lcc}
\hline\hline                                           
         $\Lambda$ (GeV)  & 0.5   & 1.0  \\        
\hline
     $\sigma(pp\rightarrow A_{D^+D^-}+\all)$ (pb)  & $1^{+7}_{-1}$  & $49^{+76}_{-33}$     \\
\hline\hline
\end{tabular*}
\end{table}

Assuming the dionium mainly decays to $D^0\bar{D}^0$, i.e., $\text{Br}[A_{D^+D^-}\rightarrow D^0\bar{D}^0]\simeq 1$, 
the cross section for the process $pp\rightarrow A_{D^+D^-}+\all\rightarrow D^0\bar{D}^0+\all$ reaches ${\cal{O}}(10 ~\text{pb})$ with the CMS kinematics in Ref.~\cite{CMS:2013fpt}, where the inclusive production cross section for the $X(3872)$ was measured to be $\sigma_{X(3872)}\text{Br}[X(3872)\rightarrow J/\psi \pi^+\pi^-]=(1.06\pm 0.11 \pm 0.15)$~nb.
The LHCb experiment has advantages in detecting the neutral $D$ mesons in comparison with the CMS. Thus, in the following, we make an estimate of the prompt inclusive production rate at LHCb.

The inclusive production cross section for the $X(3872)$ at LHCb was measured to be $\sigma_{X(3872)}\text{Br}[X(3872)\rightarrow J/\psi \pi^+\pi^-]=(5.4\pm 1.3 \pm 0.8)$~nb~\cite{LHCb:2011zzp}, which is about five times larger than that at CMS~\cite{CMS:2013fpt}.
Since the long-distance part of the production of both the $X(3872)$ and the dionium is universal at both experiments and calculable in NREFT, the difference between the productions of both states at CMS and LHCb then lies in the short-distance part, and one expects
\begin{align}
    \left(\frac{\sigma_{A_{D^+D^-}}}{\sigma_{X(3872)}}\right)_{\text{CMS}}\sim \left(\frac{\sigma_{A_{D^+D^-}}}{\sigma_{X(3872)}}\right)_{\text{LHCb}},
    \label{Eq:ratio_CMS_LHCb}
\end{align}
where $\sigma_{A_{D^+D^-}}$ and $\sigma_{X(3872)}$ are the inclusive cross sections for the productions of the dionium and the $X(3872)$, respectively.
Thus, $\sigma(pp\rightarrow A_{D^+D^-}+\all \to D^0\bar D^0 +\all)$ should be of ${\cal{O}}(0.1 ~\text{nb})$ at LHCb with the kinematics in Ref.~\cite{LHCb:2011zzp}, i.e., $\sqrt{s_{pp}}=7$~TeV, the transverse momentum $5<p_T<20$~GeV, and the rapidity $2.5<y<4.5$. 
Thus, the observation of the dionium at LHCb is promising. Next, we estimate the production rate of the dionium at PANDA, which has a high-energy resolution and thus is able to measure the properties with a better precision if the production rate is high enough.

\section{Production rate of the dionium in $p{\bar p}$ collisions}\label{Sect:production_ppbar}

\begin{figure}[tbp]
    \includegraphics[width=0.45\columnwidth]{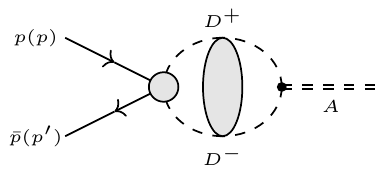}
    \caption{~The Feynman diagram for the $p\bar{p}\rightarrow  A_{D^+D^-}$ process. Here, $A$ denotes the dionium $A_{D^+D^-}$.
    Besides the strong interaction, the electric
    Coulomb interaction is present in the $D^+D^-$ loop, 
    where the analytic form of this mesons loop is shown in Eq.~\eqref{Eq:G_c}. }
    \label{fig:feynman_A_DD}
\end{figure}

\begin{figure}[tbp]
    \includegraphics[width=0.42\columnwidth]{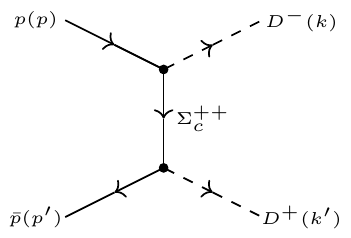}
    \caption{~The charged charm mesons production from $p\bar{p}$ with the exchange of $\Sigma_c^{++}$.}
    \label{fig:Feyn_ppb_DD}
\end{figure}

The dionium can be produced directly at the $p\bar{p}$ collisions, as shown in Fig.~\ref{fig:feynman_A_DD}. The $D^+D^-$ pair is produced at short distances in $p{\bar p}$ collisions, and the dionium is formed from the long-distance FSI of this meson pair. The short-distance process $p{\bar p}\rightarrow D^+D^-$ may be estimated by exchanging the $\Sigma_c^{++}$ and $\Sigma_c^{*++}$ in a $t$-channel scattering.
Because the coupling constant between $p$ and $\Sigma_c^* {\bar D}$ is small~\cite{Holzenkamp:1989tq,Haidenbauer:2014rva}, 
the contribution from the exchange of $\Sigma_c^{*++}$ is largely suppressed.
Therefore, we only count the contribution of the $\Sigma_c^{++}$ exchange in the process $p{\bar p}\rightarrow D^+D^-$ shown in Fig.~\ref{fig:Feyn_ppb_DD}.
Of course, there can also be exchanges of excited $\Sigma_c^{++}$ baryons. Here, we only aim at an order-of-magnitude estimate of the production cross section and thus will not consider other possible exchanges.

The effective Lagrangian free of derivative for the interaction between $p$ and $\Sigma_c {\bar D}$ can be constructed as
\begin{align}
    {\cal L}_{N\Sigma_c D}=&g_{s}\bar{\psi}_N\gamma^5\phi_D\psi_{\Sigma_c}+ \text{H.c.},
    \label{Eq:Lagrangian_BBM}
\end{align}
where $\psi_N$, $\psi_{\Sigma_c}$, and $\phi_D$ are the fields of the nucleon, the $\Sigma_c$, and the $D$ meson, respectively, and the coupling constant $g_{s}=2.69$ from the flavor SU(4) model~\cite{Haidenbauer:2010ch,Shyam:2015hqa}.
As shown in Fig.~\ref{fig:Feyn_ppb_DD}, the transition amplitude for the $p{\bar p}\rightarrow D^+D^-$ channel with the $\Sigma_c$ exchange is expressed as
\begin{align}
    i{\cal M}_{p\bar{p}\rightarrow D^+D^-}=\bar{\upsilon}(p')g_{s}\gamma^5
    \frac{iF^2_{p\Sigma_cD}}{\slashed{p}-\slashed{k}-m_{\Sigma_c}}g_{s}\gamma^5u(p),
    \label{Eq:amp_ppb_DD}
\end{align}
where $u$ and $\bar{\upsilon}$ are the spinors of the proton and antiproton, respectively; $m_{\Sigma_c}$ is the mass of $\Sigma_c$; and $F_{p\Sigma_cD}$ is the form factor. Following Refs.~\cite{Dong:2014ksa,Sakai:2020crh}, 
we take the form factor as
\begin{align}
    F^2_{p\Sigma_cD}=\frac{\Lambda_1^4}{\left[(p-k)^2-m_{\Sigma_c}^2\right]^2+\Lambda_1^4},
    \label{Eq:form_factor}
\end{align}
where $k$ is the 4-momentum of $D^-$ and the cutoff $\Lambda_1$ is taken to be $\Lambda_1=2.0~\text{GeV}$ as in Refs.~\cite{Dong:2014ksa,Sakai:2020crh}.

To estimate the production rate, we set the total initial energy of $p\bar p$ at the mass of the dionium, $\sqrt{s}=3739.3$~MeV, to maximize the cross section of $p{\bar p}\rightarrow A_{D^+D^-}$ with subsequent $A_{D^+D^-}$ decays into final states such as $D^0\bar D^0$. 
For such an initial energy, the value of $(p-k)^2/m_{\Sigma_c}^2$ is small in the center-of-mass frame, 
\begin{align}
    \frac{(p-k)^2}{m_{\Sigma_c}^2}\simeq-\frac{({\mathbf p}-{\mathbf k})^2}{m_{\Sigma_c}^2}\simeq -0.4.
\end{align}
So, one may expand the propagator in \eqref{Eq:amp_ppb_DD} in powers of $1/m_{\Sigma_c}$ and keep only the leading term for an estimate, which should suffice since we aim at an order-of-magnitude estimate.
The amplitude for $p{\bar p}\rightarrow D^+D^-$ can be reduced to
\begin{align}
    i{\cal M}'_{p\bar{p}\rightarrow D^+D^-}\simeq &-\bar{\upsilon}(p')g_{s}\gamma^5
    \frac{iF^2_{p\Sigma_cD}}{m_{\Sigma_c}}g_{s}\gamma^5u(p) + {\cal O}(m_{\Sigma_c}^{-2}).
    \label{Eq:amp_ppb_DD_reduce}
\end{align}

\begin{figure}[tb]
    \includegraphics[width=0.6\columnwidth]{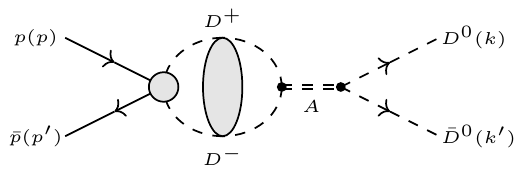}
    \caption{~The Feynman diagram for the $p\bar{p}\rightarrow  A_{D^+D^-}\rightarrow D^0{\bar D}^0$ process. $A$ denotes the dionium.}
    \label{fig:feynman_A_decay}
\end{figure}

The reduced amplitude for $p\bar{p}\rightarrow A_{D^+D^-}$ shown in 
Fig.~\ref{fig:feynman_A_DD} is
\begin{align}
    i{\cal M}_{A_{D^+D^-}}=i{\cal M}_{\text{ISI}}{\cal M}'_{p\bar{p}\rightarrow D^+D^-}\frac{{\cal N}G_c^{\Lambda}g_{\text{str}}}{4m_{D^+}m_{D^-}},
    \label{Eq:amp_total}
\end{align}
where ${\cal N}=\sqrt{8m_{A_{D^+D^-}}m_{D^+}m_{D^-}}$ accounts for the normalization of the heavy particle fields,
$G_c^{\Lambda}$ is the Coulomb propagator shown in Eq.~\eqref{Eq:G_c}, 
and $g_{\text{str}}$ is effective coupling between the dionium and $D^+D^-$ in Eq.~\eqref{Eq:coupling}. 
Here, ${\cal M}_{\text{ISI}}$ is a factor derived from the $p{\bar p}$ initial-state interaction \cite{Haidenbauer:2009ad,Dong:2014ksa,Dong:2017gaw}. The factor $\left|{\cal M}_{\text{ISI}}\right|^2$ slowly varies from 0.27 to 0.32 with the increase of $\sqrt{s}$ from 5~GeV up to 8~GeV~\cite{Dong:2014ksa}. In this work, the interesting energy region is lower, at around 3.74~GeV, and we set $\left|{\cal M}_{\text{ISI}}\right|^2\approx0.2$ for an order-of-magnitude estimation of the production rate of the dionium at PANDA.
According to the general cross section formula Eq.~\eqref{Eq:section_2_1} for the $2\rightarrow 1$ process, the cross section 
for $p{\bar p}\rightarrow A_{D^+D^-}$ is
\begin{align}
    \sigma_{A_{D^+D^-}}
    =&\frac{1}{\Gamma_{A_{D^+D^-}} \sqrt{s \lambda(s,m_p^2,m_p^2)}}
    \left|{\cal M}_{A_{D^+D^-}}\right|^2\nonumber\\
    =&\frac{(E_p^2-m_p^2)\left|{\cal M}_{\text{ISI}}\right|^2}{\Gamma_{A_{D^+D^-}} \sqrt{s \lambda( s,m_p^2,m_p^2)}} 
    \frac{m_{A_{D^+D^-}}g_{s}^4 F^4_{p\Sigma_cD}}{m_{D^+}m_{D^-}m^2_{\Sigma_c}}\left|G_c^{\Lambda}g_{\text{str}}\right|^2,
    \label{Eq:cross_section_ppb_total}
\end{align}
where $E_{p}$ and $m_p$ are the energy in the center-of-mass frame and mass of the proton, respectively; $\lambda(x,y,z)\equiv x^2+y^2+z^2-2xy-2yz-2zx$; and  $\Gamma_{A_{D^+D^-}}$ is the decay width of the dionium. 
Since the dionium mainly decays to $D^0\bar{D}^0$, i.e., $\text{Br}[A_{D^+D^-}\rightarrow D^0{\bar D}^0]\simeq 1$, 
the decay width of the dionium $\Gamma_{A_{D^+D^-}}$ is approximately equal to the partial width $\Gamma_1$ in Eq.~\eqref{Eq:Binding_energy}. Because the decay width of the dionium $\Gamma_1$ is tiny in comparison with any hadronic scale, the general $2\to1$ cross section formula Eq.~\eqref{Eq:section_2_1} is valid for $p{\bar p}\rightarrow A_{D^+D^-}$. Moreover, in Appendix \ref{Sect:one_phase_space}, we show that
\begin{align}
    \sigma_{A_{D^+D^-}}\text{Br}[A_{D^+D^-}\rightarrow D^0{\bar D}^0]=\sigma_{D^0{\bar D}^0},
\end{align}
where $\sigma_{D^0{\bar D}^0}$ is the cross section of $p\bar p\to A_{D^+D^-}\to D^0\bar D^0$ shown in Fig.~\ref{fig:feynman_A_decay}. From this relation, we confirm that 
the general cross section \eqref{Eq:section_2_1} for the  $2\rightarrow 1$ process  can be applied to $p{\bar p}\rightarrow A_{D^+D^-}$.

Results for the $p\bar{p}\rightarrow A_{D^+D^-}$ cross section, as estimated using Eq.~\eqref{Eq:cross_section_ppb_total},
at $\sqrt{s}=3739.3$ MeV with different cutoffs $\Lambda$ are demonstrated in Table \ref{Tab:cross_section_ppb}. For the cutoff being $\Lambda=0.5(1.0)$ GeV, the cross section for $p\bar{p}\rightarrow A_{D^+D^-}$, as well as for  $p\bar{p}\rightarrow A_{D^+D^-}\to D^0\bar D^0$, is $0.002(0.1)~\mu$b. 
From Ref.~\cite{PANDA:2018zjt}, the integrated luminosity of PANDA at $\sqrt{s}=3872$~MeV is about 2~fb$^{-1}$ in five months.
Assuming the same integrated luminosity at 3739.3~MeV, where the production rate is  maximal, the above estimated cross sections translate into  $0.04(2.0)\times10^8$ events.
Reconstructing the neutral $D^0$ and $\bar D^0$ mesons using the $K^\mp\pi^\pm$ mode, which has a branching fraction of about 4\%~\cite{Zyla:2020zbs}, there will still be  $\mathcal{O}(10^3\sim 10^5)$ events. Therefore, we expect that the dionium can be detected at PANDA.

\begin{table}[tb]
\caption{\label{Tab:cross_section_ppb} Order-of-magnitude estimates of the integrated cross section for $p{\bar p}\rightarrow A_{D^+D^-}$. 
In the first row, the cutoff $\Lambda$ originated from the regularized Green's function in Eq.~\eqref{Eq:G_c}.
}
\renewcommand{\arraystretch}{1.2}
\begin{tabular*}{\columnwidth}{@{\extracolsep\fill}lcc}
\hline\hline                                           
         $\Lambda$ (GeV)  & 0.5   & 1.0 \\        
\hline
     $\sigma(p\bar{p}\rightarrow A_{D^+D^-})$ ($\mu$b)  &  $0.002^{+0.013}_{-0.002}$ &  $0.1^{+0.2}_{-0.1}$ \\
\hline\hline
\end{tabular*}
\end{table}

\section{Conclusions}\label{Sect:summary}

We have predicted the binding energy and the partial width to $D^0\bar{D}^0$ of the dionium, the ground-state $D^+D^-$ hadronic atom with $J^{PC}=0^{++}$, and estimated its production cross sections (inclusive and exclusive, respectively) in $pp$ and $p{\bar p}$ colliders.

The binding energy and the decay width of the dionium were derived from the pole of the $D^0\bar{D}^0$-$D^+D^-$ coupled-channel scattering amplitude, where the potential for 
the strong interaction is a contact term derived from HQSS. In the presence of a $D^0\bar{D}^0$ bound state predicted in the LQCD calculation~\cite{Prelovsek:2020eiw}, we have predicted that the mass of the dionium $A_{D^+D^-}$  
is $(3739.3 \pm0.1)~\mathrm{MeV}$, and the decay width for $A_{D^+D^-}\rightarrow D^0{\bar D}^0$ 
is $1.8^{+1.4}_{-0.6}~\mathrm{keV}$. 
The latter is expected to be a good estimate for the total width of the ground state dionium, corresponding to a life time of order $10^{-18}$ seconds.

We have discussed the inclusive process $pp\rightarrow D^+D^-+\all$ with the MC event generator P{\scriptsize{YTHIA}}8 and
produced the dionium from the $D^+D^-$ FSI to explore an order-of-magnitude estimate of the dionium production cross section at LHC experiments.
Our result shows that the cross section for $pp\rightarrow A_{D^+D^-}+\all$ is at the $\mathcal{O}(10)$~pb level at CMS with $\sqrt{s_{pp}}=7$~TeV, which is 3 orders of magnitude
smaller than the cross section for the production of $X(3872)$ predicted in Ref.~\cite{Guo:2014sca} and measured in Ref.~\cite{CMS:2013fpt} and a few times larger at LHCb with $\sqrt{s_{pp}}=7$~TeV. 

The cross section for $p{\bar p}\rightarrow A_{D^+D^-}$ was also estimated at the energy 3739.3~MeV where the production rate should be maximized. The cross section for $p{\bar p}\rightarrow A_{D^+D^-}$ is at the $10^{-1}$ $\mu$b level, and we expect that $\mathcal{O}(10^3\sim 10^5)$ events of the reaction chain $p\bar p\to A_{D^+D^-}\to D^0\bar D^0 \to K^-\pi^+K^+\pi^-$ can be collected. A study of the hadronic atom suggested here at PANDA will be able to provide crucial information in understanding the interaction between charmed mesons and have influence in understanding the charmoniumlike states.

\begin{acknowledgments}
    Part of the numerical calculations was done at the HPC Cluster of ITP-CAS. This work is supported in part by the Chinese Academy of Sciences under Grants 
    No.~XDB34030000, No.~XDPB15, and No.~QYZDB-SSW-SYS013; by the National Natural Science Foundation of China (NSFC) under 
    Grants No.~12125507,  No.~11835015, No.~12047503, and No.~11961141012; and by the NSFC and the Deutsche Forschungsgemeinschaft 
    (DFG) through the funds provided to the TRR110 ``Symmetries and the Emergence of Structure in QCD'' (NSFC Grant No. 12070131001, DFG Project-ID 196253076).
\end{acknowledgments}

\appendix

\section{One-body phase space and the dionium line shape}\label{Sect:one_phase_space}

For the process $AB\rightarrow C$, the one-body phase space is
\begin{align}
    \int \text{d}\Pi_1=& \int \frac{dp^3_C}{(2\pi)^32E_C}(2\pi)^4\delta^4(p_A+p_B-p_C)\nonumber\\
    =& \frac{2\pi \delta(E_A+E_B-E_C)}{2(E_A+E_B)}.
    \label{Eq:phase_space_ori}
\end{align}
The Dirac $\delta$ function in Eq.~\eqref{Eq:phase_space_ori} can be approximately rewritten as
\begin{align}
    \delta(\sqrt{s}-E_C)=&\frac{1}{i\pi} \lim_{\Gamma\to 0}\left(\frac{1}{\sqrt{s}-E_C-i\Gamma/2}\right)\nonumber\\
    \simeq& \frac{1}{i\pi}\left[\frac{\sqrt{s}-E_C+i\Gamma/2}{(\sqrt{s}-E_C)^2+\Gamma^2/4}\right]_{\sqrt{s}\rightarrow E_C}\nonumber\\
    =&\frac{2}{\pi\Gamma},
    \label{Eq:delta_deal}
\end{align}
with $\Gamma$ the decay width of the $C$ particle, which should be small for the approximation to work. The one-body phase space with the narrow-width approximation 
for the final state can be rewritten as
\begin{align}
    \int \text{d}\Pi_1=\frac{2}{\Gamma(E_A+E_B)}.
    \label{Eq:one_body_phase_space}
\end{align}

Applying the narrow-width approximation, the cross section of $AB\rightarrow C$ in the center-of-mass frame is
\begin{widetext}
\begin{align}
    \sigma_{C}=&\frac{1}{2E_A2E_B\left|\upsilon_A-\upsilon_B\right|} \int \frac{dp^3_C}{(2\pi)^32E_C}
            \left|{\cal M}_{AB\rightarrow C}\right|^2(2\pi)^4\delta^4(p_A+p_B-p_C)\nonumber\\
    =&\frac{1}{4\left|E_B {\mathbf p}+E_A {\mathbf p}\right|} \frac{2\pi \delta(E_A+E_B-E_C)}{2(E_A+E_B)}
    \left|{\cal M}_{AB\rightarrow C}\right|^2\nonumber\\
    \simeq&\frac{1}{\Gamma \sqrt{s \lambda( s,m_A^2,m_B^2)}}\left|{\cal M}_{AB\rightarrow C}\right|^2,
    \label{Eq:section_2_1}
\end{align}
\end{widetext}
where $s=(E_A+E_B)^2$ and $\mathbf{p}$ is
the 3-momentum of initial particles in the center-of-mass frame,  
$|{\mathbf p}| = {\sqrt{\lambda( s,m_A^2,m_B^2)}}/{(2\sqrt{s})}$.
Equation~\eqref{Eq:section_2_1} can be used to calculate the cross section of the $2\to1$ process $AB\rightarrow C$ at $\sqrt{s}=m_C$
for a narrow $\Gamma$.

For $\sqrt{s}>m_C$ or a large decay width of $C$, one needs to consider the cross section for $p{\bar p}\rightarrow A_{D^+D^-}\rightarrow D^0{\bar D}^0$ shown in 
Fig.~\ref{fig:feynman_A_decay}. The amplitude for this process is
\begin{align}
    i{\cal M}_{D^0{\bar D}^0}= i{{\cal M}_{\text{ISI}}}{\cal M}'_{p\bar{p}\rightarrow D^+D^-}\frac{{\cal N}'G_c^{\Lambda}g_{\text{str}}}{2m_{D^+}2m_{D^-}} \frac{g_{\text{str}}}{2m_{A_{D^+D^-}}(\sqrt{s}-m_{A_{D^+D^-}}+i\Gamma_{A_{D^+D^-}}/2)},
    \label{Eq:amp_total_app}
\end{align}
where ${\cal N}'=\sqrt{8m_{A_{D^+D^-}}m_{D^+}m_{D^-}}\sqrt{8m_{A_{D^+D^-}}m_{D^0}m_{\bar{D}^0}}$ accounts for the normalization 
of the heavy particle fields and ${\cal M}'_{p\bar{p}\rightarrow D^+D^-}$ is the amplitude shown 
in Eq.~\eqref{Eq:amp_ppb_DD_reduce}. The Coulomb propagator $G_c^{\Lambda}$ and the effective coupling constant $g_{\text{str}}$ are given in Eqs.~\eqref{Eq:G_c}
and \eqref{Eq:coupling}, respectively. The cross section for $p{\bar p}\rightarrow A_{D^+D^-}\rightarrow D^0{\bar D}^0$ is
\begin{widetext}
\begin{align}
\label{Eq:cross_section_ppb}
    \sigma_{D^0{\bar D}^0}&=\frac{1}{4}\frac{1}{4E_pE_{\bar p}\left|v_p-v_{\bar p}\right|}\int \text{d}\Omega \frac{{|\mathbf k}|}{(2\pi)^24\sqrt{s}}
    \left|{\cal M}_{D^0{\bar D}^0}\right|^2\\
    &=\frac{(E_p^2-m_p^2)}{8\pi s}{\left|{\cal M}_{\text{ISI}}\right|^2}\sqrt{\frac{\lambda(s,m_{D^0}^2,m_{\bar{D}^0}^2)}{\lambda(s,m_p^2,m_{\bar p}^2)}}
    \frac{m_{D^0}m_{{\bar D}^0}g_{s}^4 F^4_{p\Sigma_cD}}{m_{D^+}m_{D^-}m^2_{\Sigma_c}}
   \frac{ \left|G_c^{\Lambda}g^2_{\text{str}}\right|^2}{(\sqrt{s}-m_{A_{D^+D^-}})^2+\Gamma_{A_{D^+D^-}}^2/4}. \nonumber
\end{align}
\end{widetext}

In the center-of-mass frame, one has $s=(E_p+E_{\bar p})^2=(E_{D^0}+E_{\bar{D}^0})^2$ and 
$p\cdot p'=p^0p'^0-{\mathbf p}\cdot{\mathbf p}'=E_p^2+(E_p^2-m_p^2)=2E_p^2-m_p^2$, 
with $E_{p}$ ($E_{\bar{p}}$) and $m_p$ ($m_{\bar{p}}$) the energy and mass of the proton (antiproton), respectively. 

At LO, the decay width for $A_{D^+D^-}\rightarrow D^0{\bar D}^0$ is   
\begin{align}
    \Gamma_{D^0{\bar D}^0}\simeq &\frac{1}{2m_{A_{D^+D^-}}}\frac{|\mathbf{k}|\left|g_{\text{str}}\right|^2}{4\pi m_{A_{D^+D^-}}}
    8m_{A_{D^+D^-}}m_{D^0}m_{\bar{D}^0}\nonumber\\
    =&\frac{m_{D^0}m_{\bar{D}^0}\left|g_{\text{str}}\right|^2}{\pi m_{A_{D^+D^-}}}
    \frac{\sqrt{\lambda(s,m_{D^0}^2,m_{\bar{D}^0}^2)}}{2\sqrt{s}},
    \label{Eq:decay_width_D0D0}
\end{align}
and the cross section for 
$p{\bar p}\rightarrow A_{D^+D^-}\rightarrow D^0{\bar D}^0$ at $\sqrt{s}=m_{A_{D^+D^-}}$ can be written as
\begin{widetext}
    \begin{align}
        \sigma_{D^0{\bar D}^0}\simeq &\frac{(E_p^2-m_p^2)}{4 \sqrt{s \lambda(s,m_p^2,m_{\bar p}^2)}}{\left|{\cal M}_{\text{ISI}}\right|^2}
        \frac{m_{A_{D^+D^-}}g_{s}^4 F^4_{p\Sigma_cD}}{m_{D^+}m_{D^-}m^2_{\Sigma_c}}
        \left|G_c^{\Lambda}g_{\text{str}}\right|^2\frac{\Gamma_{A_{D^+D^-}}}{(\sqrt{s}-m_{A_{D^+D^-}})^2+\Gamma_{A_{D^+D^-}}^2/4}
        \frac{\Gamma_{1}}{\Gamma_{A_{D^+D^-}}}\nonumber\\
        =&\sigma_{A_{D^+D-}}\mathrm{Br}[A_{D^+D^-}\rightarrow D^0\bar{D}^0],
    \end{align}
\end{widetext}
which indicates that at $\sqrt{s}=m_{A_{D^+D^-}}$ the cross section $\sigma_{A_{D^+D^-}}$ for $p{\bar p}\rightarrow A_{D^+D^-}$ 
in Eq.~\eqref{Eq:cross_section_ppb_total} is equal to 
$\sigma_{D^0{\bar D}^0}/\mathrm{Br}[A_{D^+D^-}\rightarrow D^0\bar{D}^0]$. Moreover, the $\sigma_{D^0{\bar D}^0}$ computed
by Eq.~\eqref{Eq:cross_section_ppb} can be applied to explore the line shape of $A_{D^+D^-}$ in the $D^0{\bar D}^0$ 
invariant-mass distribution, while Eq.~\eqref{Eq:section_2_1} just estimates the cross section at 
$\sqrt{s}=m_{A_{D^+D^-}}$.

\bibliography{production_Dm_Dp_ref}



\end{document}